# The Global Food Trade Network as a Complex Adaptive System: A Review of Structure, Evolution, and Resilience

Zebiao Li [a], Xueying Wu [b], Chengyi Tu [b, *]

[a] Keyi College of Zhejiang Sci-Tech University; Shaoxing, 312369, China

[b] School of Economics and Management, Zhejiang Sci-Tech University; Hangzhou, 310018, China

[*] Corresponding author. Email: chengyitu1986@gmail.com

## Abstract

The global food system has metamorphosed from a loose aggregation of bilateral exchanges into a highly intricate, interdependent Global Food Trade Network (FTN). This comprehensive review synthesizes the extant literature to examine the FTN through the rigorous lens of complex network science, moving beyond traditional economic trade models to quantify the system's topological architecture. We delineate the network's historical transition from a unipolar, efficiency-driven system dominated by Western hegemony to a multipolar, regionalized structure characterized by high clustering and scale-free heterogeneity. Special emphasis is placed on the dual nature of connectivity, which functions simultaneously as a buffer against local production variances and a conduit for global contagion. By conceptualizing the FTN as a multiplex system—distinguishing between the robust topology of wheat, the brittle regionalism of rice, and the polarized "dumbbell" structure of soy—we elucidate the distinct structural vulnerabilities inherent in modern food security. Furthermore, we analyze the impact of recent high-magnitude shocks, specifically the COVID-19 pandemic and the Russia-Ukraine conflict, illustrating the critical trade-off between logistical efficiency and systemic resilience. The review concludes by assessing the future trajectory of the network under anthropogenic climate change, predicting a poleward migration of comparative advantage that necessitates a paradigm shift from isolationist protectionism to cooperative network redundancy.

**Keywords**: Global Food Trade Network (FTN); Complex Network Analysis; Food Security; Systemic Resilience; Multiplex Networks; Food-Energy-Water Nexus; Climate Adaptation

# 1 Introduction

The global food system, once conceptualized as a loose aggregation of bilateral exchanges governed by the invisible hand of comparative advantage, has metamorphosed into a highly intricate, interdependent web of international trade[1]. This system now facilitates the massive redistribution of calories, nutrients, and virtual water required to sustain a growing human population that has long surpassed the carrying capacity of local agricultural systems in many regions[2,3]. As nations increasingly rely on imports to bridge the widening gap between domestic production and consumption, the Global Food Trade Network (FTN) has emerged not merely as a market mechanism, but as a critical determinant of planetary food security and a distinct complex adaptive system[4-6].

For much of the 20th century, traditional economic analyses approached international trade through the reductionist frameworks of Ricardian comparative advantage, Heckscher-Ohlin factor endowments, and gravity models[7-11]. These methodologies, while powerful in explaining the drivers of trade volume and price formation between dyads, treated trade relationships as independent events. They operated under the assumption of equilibrium, largely ignoring the higher-order interactions, feedback loops, and emergent structural properties that define complex systems[12-15]. By viewing the world as a collection of isolated trading pairs, these traditional models failed to capture the systemic risks and structural dependencies that allow a localized shock—such as a drought in the Pontic Steppe or a port strike in Rotterdam—to propagate rapidly across the entire globe, precipitating cascading failures in remote, seemingly unconnected markets[4,16,17].

The "network turn" in agricultural economics represents a paradigm shift, moving the focus from the volume of trade to the structure of trade[18,19]. Complex Network Analysis (CNA) offers the robust theoretical framework necessary for this shift. By abstracting countries as nodes and trade flows as edges, CNA enables researchers to quantify the topological architecture of the global food system. It moves beyond simple trade balances to identify key players based on their structural position (centrality), uncover hidden vulnerabilities through community detection, and map the specific mechanisms of shock propagation through connectivity pathways[20-23].

This comprehensive review synthesizes the extant literature, examining the FTN through the rigorous lens of complex network science. We provide an exhaustive analysis of the network's topological architecture, delineating its historical transition from a unipolar, efficiency-driven system dominated by Western hegemony to a multipolar, regionalized structure characterized by high clustering and scale-free heterogeneity. Special emphasis is placed on the dual nature of connectivity, which functions simultaneously as a buffer against local production variances—pooling risk across the globe—and a conduit for global contagion. Through a detailed examination of empirical studies, methodological advancements, and recent systemic shocks—including the COVID-19 pandemic and the Russia-Ukraine conflict—we elucidate the structural vulnerabilities inherent in the modern food system and offer insights into the resilience mechanisms requisite for navigating future climate-induced disruptions[24-27].

# 2 Methodological Foundations of Network Construction and Metrics

The application of complex network science to the global food system reveals a hierarchical architecture defined by distinct properties at the micro-, meso-, and macro-scales. As illustrated in Figure 1, the transformation of raw bilateral trade data into a topological formalism allows for the quantification of systemic stability and the identification of critical vulnerabilities.

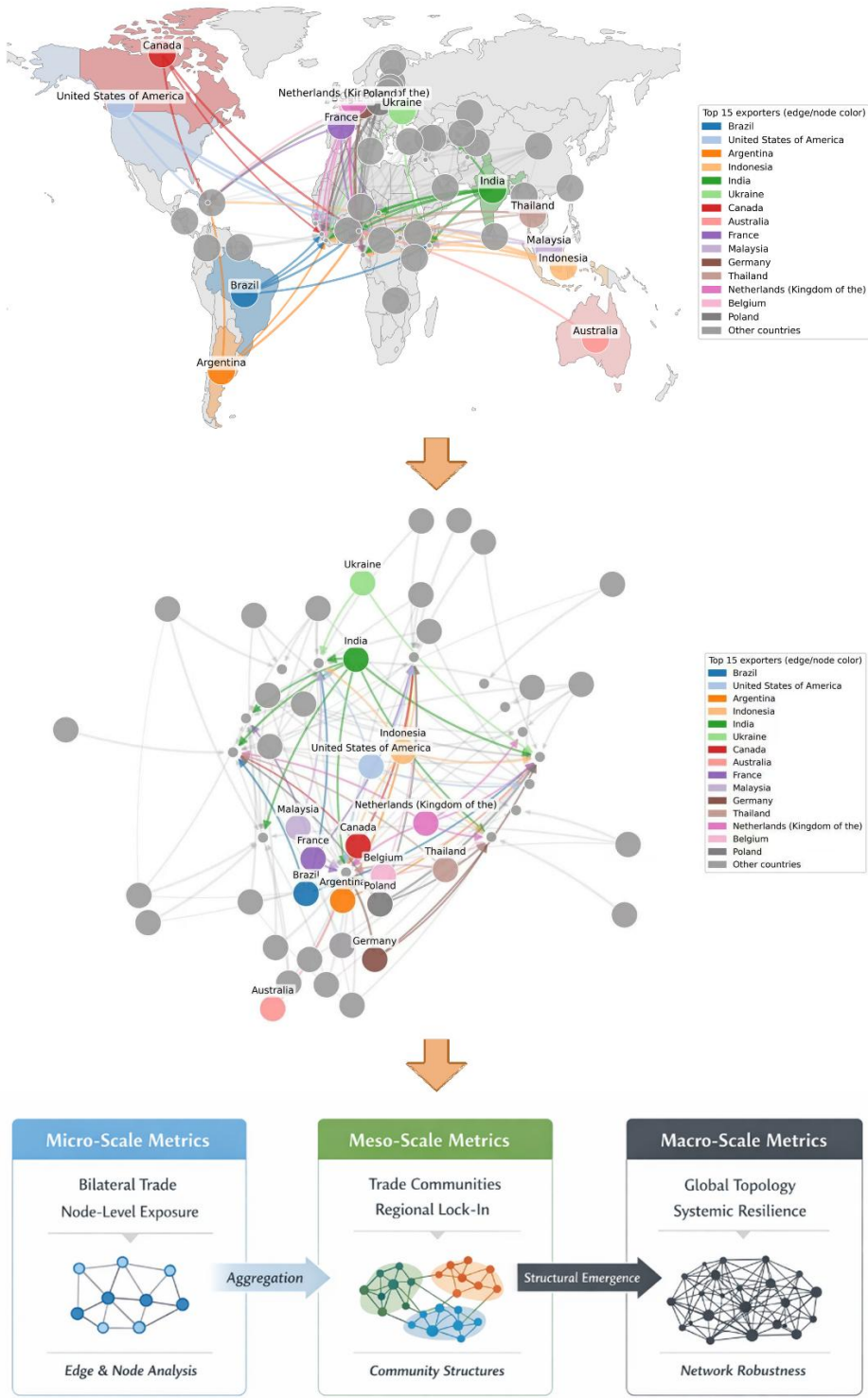

**Figure 1. Methodological framework for the analysis of the Global Food Trade Network (FTN).** The schematic illustrates the translation of geospatial trade data into a complex network model and the subsequent multi-level analysis. The geospatial representation of the global food system, mapping bilateral trade flows (edges) between countries (nodes), highlighting the dominance of top exporters such as Brazil, the United States, and Russia. The topological abstraction of the FTN as a complex network graph. This visualization removes geographic constraints to reveal structural clustering, hub-and-spoke formations, and the density of inter-regional connections. The hierarchical metrics framework used to quantify network properties.

## 2.1 Topological Abstraction and Network Construction

The application of complex network theory to the global food system is predicated on a rigorous mathematical formalization that translates heterogeneous trade data into a computable topological structure. Formally, the Global Food Trade Network (FTN) is defined as a graph $G = (N, E, W)$, where $N$ denotes the set of nodes (countries or customs territories), $E$ represents the set of edges (trade flows), and $W$ captures the weights or intensity of these interactions. This formalism provides the analytical substrate for quantifying systemic properties ranging from local influence to global resilience[28,29]. Furthermore, temporal resolution plays a vital role in network construction[30,31]. While trade flows are continuous, data is typically aggregated on an annual basis. Consequently, longitudinal studies often span extensive periods (e.g., 1986–2022) to capture evolutionary trends, whereas higher-frequency analyses (monthly or quarterly) are increasingly utilized to characterize the propagation of rapid shocks, such as the 2008 food price crisis or the onset of the COVID-19 pandemic.

In the context of the FTN, the set of nodes ($N$) invariably comprise the 190+ countries and independent customs territories reporting trade data to international bodies such as the United Nations Comtrade database or the Food and Agriculture Organization (FAOSTAT)[32]. While the definition of a node is relatively static, the definition of an edge ($E$) requires careful consideration of directionality.

Trade flows are inherently asymmetrical. An export of wheat from the United States (node $i$) to Egypt (node $j$) establishes a flow of physical goods in one direction and a flow of financial capital in the reverse. Consequently, most rigorous studies model the FTN as a directed network (digraph)[5,20,33]. This directionality is not a trivial detail; it is crucial for analyzing dependencies and power dynamics.

Perhaps the most critical methodological decision in constructing the FTN is the choice of weighting mechanism ($W$). The weight assigned to an edge determines the "cost" of traversal and the "strength" of connections, fundamentally altering the network's topology and the resulting interpretation of "importance" or "centrality."[34,35] The literature presents divergent findings based on five primary weighting approaches (see Table 1):

Binary (Unweighted) Topology[12]: In a binary network, a link is assigned a value of $w_{ij} = 1$ if any trade exists, and 0 otherwise. This approach, while seemingly simplistic, is instrumental in analyzing the pure connectivity structure of the system. It reveals the potential pathways for the diffusion of information, policy norms, or biological pathogens (e.g., pests, foot-and-mouth disease), independent of the volume of trade. A single kilogram of contaminated meat can transmit a virus just as effectively as a thousand tons; thus, binary topology is essential for biosecurity analysis.

Physical Volume (Mass)[23]: Weights represent the physical tonnage ($kg$ or $MT$) of food traded. This metric is the standard for analyzing logistics, shipping, infrastructure requirements, and the physical burden on the planet's transport systems. However, mass-based weighting can be misleading when aggregating diverse commodities. Comparing a ton of staple wheat (essential for survival) to a ton of high-value spices, wine, or saffron distorts the food security implications. Mass emphasizes bulk commodities like grains and oilseeds while marginalizing nutritionally dense but lower-volume goods.

Monetary Value[18]: Weights represent the economic value, typically in US Dollars. This is the preferred metric for economists analyzing value chains, balance of payments, and economic dependency. However, monetary weights are highly susceptible to distortion by price fluctuations, inflation, and exchange rate volatility. A global price spike can artificially inflate the "weight" of the network without any increase in the physical transfer of food. Moreover, high-value luxury goods (e.g., chocolates, spirits) can dominate low-value staples in a monetary network, obscuring the flows that actually prevent famine.

Caloric and Nutritional Content[29]: A growing body of literature argues for weighting edges by caloric energy ($kcal$) or specific nutrients (protein, fat). This approach offers the most holistic view of food security, normalizing different commodities into a single metric of human sustenance. Studies utilizing caloric weights highlight the critical role of staple grains (wheat, rice, maize) over high-value perishables in maintaining global survival. Recent findings indicate a significant divergence in centrality rankings when switching from monetary to caloric weights; nations like Ukraine and Argentina, which export massive volumes of low-cost calories, rise in prominence, while nations exporting high-value processed foods (e.g., Netherlands, Germany) may appear less central to basic survival.

Virtual Water[36]: Advanced environmental analyses weigh edges by the "virtual water" embedded in the production of the traded food. This links the trade network to the global hydrological cycle, revealing how water-scarce nations (e.g., in the Middle East) effectively "import" water resources in the form of grain from water-abundant nations. This perspective transforms the FTN into a mechanism for global water redistribution.

**Table 1: Comparative analysis of weighting regimes in food trade networks.**

| Weighting Regime (W) | Unit | Analytical Focus | Limitations |
|---|---|---|---|
| Binary | (0,1) | Pathway existence; diffusion of | Ignores flow intensity; a marginal link is |

| Topology | | pests, pathogens, and information. | equal to a major artery. |
|---|---|---|---|
| Physical Mass | Metric Tons ($MT$) | Transport burden; port capacity; physical food availability. | Distorts value; conflates low-value staples (wheat) with high-value/low-volume goods (saffron). |
| Monetary Value | USD ($) | Value chains; balance of payments; economic dependency. | Sensitive to inflation, price spikes, and exchange rates; obscures caloric security. |
| Nutritional Content | $kcal$, Protein ($g$) | Human sustenance; carrying capacity; nutritional resilience. | Requires conversion factors; masks economic value; complicates non-edible agricultural trade. |
| Virtual Resources | $m^3$ water, $ha$ land | Environmental footprint; virtual water trade (VWT); land displacement. | Indirect metric; relies on crop-specific water/land footprint models. |

## 2.2 Micro-Scale Metrics

Micro-scale analysis focuses on the positioning and power of individual nodes within the global aggregate[37]. As summarized in Table 2, a nation's influence is multi-dimensional, capable of being characterized by its volume (Strength), its connectivity (Degree), or its strategic position as a bridge (Betweenness)[5].

Capacity (Degree, Strength)[18,20,38]: Node strength ($s_i$) quantifies the total volume of flow, identifying the "heavyweights" of the system. However, distinguishing between in-degree and out-degree is crucial for assessing resilience. A high $k_{in}$ suggests a diversified import portfolio resilient to partner-specific shocks, whereas high $k_{out}$ indicates systemic influence.

Control (Betweenness, Closeness)[16,39]: Betweenness Centrality ($BC_i$) identifies "gatekeeper" nodes that control the shortest paths between other pairs of nations. These nodes may not always be the largest producers, but they possess disproportionate power to disrupt global flows, acting as potential bottlenecks in the event of logistical or geopolitical crises.

Prestige (PageRank, Eigenvector)[40-42]: Algorithms like PageRank ($P_i$) offer a recursive measure of influence, where a node's importance is derived from the importance of its neighbors. This is particularly useful for identifying the core members of the "Rich Club" who trade primarily with other influential nations.

**Table 2: Micro-scale metrics to quantify the influence, capacity, and position of individual nations.**

| Metric | Symbol | Definition | Interpretation in Food Trade Network |
|---|---|---|---|
| In-Degree | $k_i^{in}$ | $\sum_j a_{ji}$ | The number of unique suppliers a country has. High $k^{in}$ implies resilience against partner-specific shocks. |

| Out-Degree | $k_i^{out}$ | $\sum_j a_{ij}$ | The number of unique export destinations. High $k^{out}$ implies low dependence on any single buyer (demand-side resilience). |
|---|---|---|---|
| Node Strength | $s_i$ | $s_i^{in} = \sum_j w_{ji}$<br>$s_i^{out} = \sum_j w_{ij}$ | The total flux (mass, calories, or USD) entering or leaving a node. Distinguishes "major players" from "minor players" regardless of connection count. |
| Betweenness Centrality | $BC_i$ | $\sum_{s \neq i \neq t} \frac{\sigma_{st}(i)}{\sigma_{st}}$ | The fraction of shortest paths passing through node $i$. Identifies logistical hubs, chokepoints, and "gatekeepers" (e.g., Netherlands, Turkey). |
| Closeness Centrality | $CC_i$ | $\frac{N-1}{\sum_{j \neq i} d_{ij}}$ | How quickly a node can reach (or be reached by) all other nodes. High $CC_i$ implies rapid transmission of price shocks or pathogens. |
| PageRank / Eigenvector | $P_i$ / $x_i$ | $Ax = \lambda x$ | Measures influence based on the importance of neighbors. A country has high rank if it trades with other high-rank heavyweights (e.g., USA-China nexus). |

## 2.3 Meso-Scale Metrics

Moving beyond individual nations, meso-scale metrics reveal the modular architecture of the FTN—how nodes cluster into functional trading blocs[22]. As outlined in Table 3, this level of analysis highlights the tension between globalization and regionalization.

Compartmentalization (Modularity, Assortativity)[4,43,44]: The FTN exhibits significant modularity, indicating that the world is partitioned into distinct communities. While high modularity can act as a "firebreak" to contain localized contagions, it also suggests that peripheral nations are "locked in" to specific regional hegemons, limiting their ability to switch partners during crises.

Cohesion (Clustering)[25,45]: The high clustering coefficient ($C_i$) reflects the prevalence of transitivity in trade (triadic closure). This is driven by geographical proximity and Regional Trade Agreements, creating tight-knit local cliques.

Stratification (Rich-Club & Core-Periphery)[46,47]: Despite the network generally being disassortative (large hubs feeding small nodes), the core exhibits a "Rich-Club" phenomenon. The top economies form a dense, exclusive clique that controls the vast majority of global reserves, creating a structural core-periphery divide that defines global inequality.

Table 3: Meso-scale metrics to quantifying the modular structure, regionalization, and mixing patterns.

| Metric | Symbol | Definition | Interpretation in Food Trade Network |
|---|---|---|---|
| Clustering Coefficient | $C_i$ | $\dfrac{2t_i}{k_i(k_i-1)}$ | Probability that two trade partners of country $i$ also trade with each other. High $C$ indicates tight regional trade blocs (e.g., EU, ASEAN). |
| Modularity | $Q$ | $\dfrac{1}{2m}\sum_{ij}(A_{ij}-\dfrac{k_i k_j}{2m})\delta(c_i,c_j)$ | Measures the strength of division into modules. High $Q$ suggests a fragmented world of distinct trading blocs; low $Q$ suggests global integration. |
| Assortativity | $r$ | Pearson correlation of degrees | The preference for nodes to attach to similar nodes. The FTN is disassortative ($r<0$), meaning hubs (major exporters) tend to feed peripheral small nodes. |
| Rich-Club Coefficient | $\phi(k)$ | $\dfrac{2E_{>k}}{N_{>k}(N_{>k}-1)}$ | The density of connections among high-degree nodes. A "Rich Club" exists if $\phi(k)$ is high, indicating the major powers form a dense, exclusive clique. |
| Coreness (k-shell)[48] | $k_s$ | Recursive pruning of nodes with degree $<k$ | Decomposes the network into layers. High $k$-shell nodes form the nucleus of the global system; low $k$-shell nodes are the dependent periphery. |

## 2.4 Macro-Scale Metrics

Macro-scale metrics characterize the global topology of the FTN, providing a synoptic view of how the system balances integration, efficiency, and heterogeneity[5,20]. These measures reveal the fundamental structural constraints governing global food security (see Table 4).

Saturation and Connectedness (Density)[18,20]: The most fundamental measure of system coherence is Network Density ($\rho$), defined as the ratio of existing trade links to the total number of possible links in the graph. In the context of the FTN, $\rho$ serves as a metric of saturation.

Integration and Systemic Risk (Path Length and Diameter)[18,20,49]: Global Integration $L$ measures the average number of steps (intermediary trade partners) required to connect any two nodes. The FTN consistently exhibits a low $L$, a hallmark of "Small-World" networks. The Diameter $D$ represents the longest shortest path in the network, defining the maximum topological distance between any two nations. A shrinking $D$ over time indicates that the "trade world" is becoming smaller, bringing peripheral nations closer to the core.

Transmission Capability (Global Efficiency)[4]: While related to path length, Global Efficiency $E_{glob}$

provides a more mathematically robust measure of the system's parallel processing capability. Defined as the average of inverse shortest path lengths, $E_{glob}$ quantifies how effectively the network exchanges information and physical goods.

Heterogeneity and Inequality (Power-Law Exponent)[4,12,18]: The structural hierarchy of the FTN is captured by the Power-Law Exponent ($\gamma$) of the degree distribution ($P(k) \sim k^{-\gamma}$). Empirical analysis places the FTN's exponent in the range of $2 < \gamma < 3$, classifying it as a scale-free network. This metric quantifies extreme heterogeneity. It confirms a system dominated by a few "super-hubs" (high-degree nodes like the USA, Brazil, China) amidst a vast sea of poorly connected peripheral nations.

**Table 4: Macro-scale metrics to Quantify the global architecture, efficiency, and robustness.**

| Metric | Symbol | Definition | Interpretation in Food Trade Network |
| --- | --- | --- | --- |
| Network Density | $\rho$ | $\dfrac{E}{N(N-1)}$ | The ratio of existing links to possible links. Indicates the overall "connectedness" of the global system. |
| Average Path Length | $L$ | $\dfrac{1}{N(N-1)}\sum_{i \neq j} d_{ij}$ | The average number of steps between any two countries. Low $L$ (Small-World property) facilitates efficient distribution but rapid shock propagation. |
| Global Efficiency | $E_{glob}$ | $\dfrac{1}{N(N-1)}\sum_{i \neq j} \dfrac{1}{d_{ij}}$ | A measure of transmission capability. Unlike $L$, it is robust to disconnected components. Measures how well the network distributes food globally. |
| Diameter | $D$ | $\max(d_{ij})$ | The longest shortest path in the network. Defines the "size" of the trade world in topological steps. |
| Power-Law Exponent | $\gamma$ | $P(k) \sim k^{-\gamma}$ | Characterizes the scale-free nature. For FTN, $2 < \gamma < 3$. Indicates extreme inequality: few hubs, many peripherals. |

# 3 Structural Properties of Global Food Trade Network

The topological structure of the FTN reveals profound insights into the stability and efficiency of the global food system. Empirical analyses across multiple decades consistently identify specific non-random structural characteristics: the "Small-World" phenomenon, the "Scale-Free" degree distribution, and distinct core-periphery architectures[5,20].

## 3.1 Small-World Phenomenon and Global Integration

The "Small-World" property, first formalized by Watts and Strogatz, describes networks that exhibit both high local clustering and short average path lengths[49]. Empirical studies consistently confirm that the FTN is a quintessential small-world network[23,25].

The clustering coefficient of the global food trade network is typically significantly higher than that of a comparable random graph ($C_{FTN} \gg C_{rand}$). This high clustering is driven by two primary factors: geographical proximity and political integration. Countries tend to form tightly connected "cliques" within their regions (e.g., Europe, South America, Southeast Asia) due to lower transport costs and Regional Trade Agreements (RTAs) such as the EU Single Market, NAFTA (now USMCA), and ASEAN[9,45].

Despite this intense regional clustering, the network maintains a remarkably short Average Shortest Path Length ($L$), typically ranging between 1.8 and 2.5 steps. This implies that, on average, any two countries in the world are connected by fewer than three trade links. This "short-circuiting" of the network is facilitated by global hubs—major exporters like the United States, Brazil, and Russia—that maintain connections across disparate regional clusters[4,12,18]. These hubs act as "shortcuts" in the graph, linking the South American soybean cluster to the East Asian consumption cluster, or the Black Sea wheat cluster to the North African import cluster.

This architecture presents a fundamental trade-off. While it maximizes distributive efficiency, allowing production deficits in one region to be compensated rapidly by distant surpluses, it simultaneously creates a "hyper-connected" substrate where local perturbations—such as price spikes, export bans, or pathogen outbreaks—can cascade globally with minimal latency.

## 3.2 Scale-Free Heterogeneity and the Power Law

The degree distribution $P(k)$ of the FTN—the probability that a randomly selected country has $k$ trade partners—follows a power-law distribution ($P(k) \sim k^{-\gamma}$), classifying it as a "scale-free" network[12,50].

Analyses plotting the cumulative degree distribution on a log-log scale reveal a distinct linear relationship in the tail, a signature of the power law. The exponent $\gamma$ typically falls between 2 and 3, indicating a system with extreme heterogeneity. This distribution signifies that the vast majority of nodes (the "periphery") have very few trade connections, often relying on one or two partners for their food security. Conversely, a tiny minority of "hub" nodes (the "core") possess a disproportionately large number of links, often numbering in the hundreds.

This "hub-and-spoke" architecture dictates the network's robustness profile. Scale-free networks are demonstrably robust to random failures[16]. If a random node—likely a poorly connected peripheral country like Nepal or Bolivia—suffers a harvest failure or ceases trading, the global connectivity of the network remains virtually unchanged. However, the network is extremely fragile to targeted attacks or the failure of hub nodes. The removal or isolation of a major hub, such as the United States (due to a trade war) or Russia (due to sanctions), can cause the entire network to fragment into disconnected components, precipitating a systemic collapse in global food availability.

## 3.3 Assortativity and the Core-Periphery Structure

Beyond the foundational small-world and scale-free properties, recent literature has focused on advanced metrics that provide deeper insight into the network's hierarchy and mixing patterns.

Assortativity ($r$) measures the preference of nodes to attach to others with similar degrees. The FTN generally exhibits disassortativity ($r < 0$), meaning that high-degree nodes (global hubs) tend to connect with low-degree nodes (peripheral importers)[44]. This negative correlation confirms the "feeder" role of major exporters: a few massive hubs distribute food to hundreds of smaller nations that are otherwise unconnected. This structure maximizes global coverage but deepens the dependency of the periphery on the core.

Despite the general disassortativity, the FTN exhibits a pronounced "Rich-Club" phenomenon among its very top nodes. The Rich-Club Coefficient $\phi(k)$ quantifies the density of connections strictly among nodes with a degree greater than $k$. Empirical analysis reveals that the world's top economies (e.g., USA, China, Germany, Netherlands, Brazil) form a tightly knit clique where almost every member trades with every other member[47]. This "club" controls the vast majority of global grain reserves and trade routes, creating a two-tier system: a highly integrated, resilient core and a dependent, fragile periphery. The core nations enjoy redundant supply chains and high resilience, while peripheral nations are tethered to the core with little horizontal connectivity among themselves.

Advanced algorithms, specifically k-shell decomposition, allow researchers to partition the network into a dense "core" and a sparse "periphery" based on recursive connectivity[48]. This analysis confirms that the "Rich Club" is not just an artifact of high degree, but a structural reality. The core is composed of nations that trade intensively with each other and with the periphery, while the periphery consists of nations that trade almost exclusively with the core and rarely with each other. This structural feature explains why South-South trade (direct trade between developing nations) remains underdeveloped compared to North-South or North-North trade, although recent trends suggest a gradual thickening of peripheral connections as emerging economies seek to bypass traditional Western hubs.

# 4 Network Centrality and Stratification of Global Trade

In the context of the Global Food Trade Network (FTN), structural position translates directly into geopolitical leverage. Centrality metrics serve as quantitative proxies for this power, allowing researchers to move beyond aggregate trade volumes to identify the distinct architectures of influence[5]. The literature differentiates between three distinct modes of influence: Capacity (Degree/Strength), Brokerage (Betweenness), and Prestige (Eigenvector/PageRank).

## 4.1 Degree and Strength Centrality

Degree centrality ($k$) and Node Strength ($s$) quantify the sheer magnitude of a nation's participation in the global system[20,38]. These metrics distinguish the primary sources and sinks of the global calorie flow.

Empirical analysis consistently identifies a stable oligopoly of exporters—principally the United States, Brazil, Russia, and arguably India—dominating the out-strength ($s_{out}$) rankings[1,25]. These nations constitute the "source" nodes of the global caloric river. A high out-degree ($k_{out}$) for these actors signifies market diversification; by exporting to a vast array of partners, they insulate themselves against demand shocks or political fallout from any single buyer, thereby enhancing their own structural resilience[6].

Conversely, in-strength ($s_{in}$) highlights the dominant consumers. China and Japan routinely anchor this metric, reflecting massive import requirements for soy and feed grains. However, a critical distinction must be drawn between Terminal Sinks and Intermediary Hubs. While China acts as a terminal sink (consuming what it imports), nations like the Netherlands and Belgium exhibit high values for both $s_{in}$ and $s_{out}$. This characterizes them not as primary producers, but as global entrepôts—processing and re-export centers that leverage logistical efficiency rather than agricultural land area[51].

## 4.2 Betweenness Centrality

While strength measures volume, Betweenness Centrality ($BC$) measures control. Defined as the frequency with which a node appears on the shortest paths between other pairs of nodes, $BC$ identifies the system's topological "gatekeepers."[39]

Nations with high betweenness—notably the United States, the Netherlands, France, and strategically located actors like Turkey and Egypt—act as bridges connecting disparate trade communities[35]. For example, Turkey acts as a crucial topological bridge between the Black Sea production sphere (Russia/Ukraine) and the consumption clusters of the Middle East and North Africa.

High-betweenness nodes represent the system's most acute points of failure. Unlike a peripheral failure, the removal or disruption of a high-$BC$ node forces trade flows to reroute through longer, less efficient, and more costly pathways[4,16]. The "Rotterdam Effect" illustrates this: a strike or blockade at Dutch ports would not only affect Dutch trade but would sever the most efficient entry point for the entire European hinterland. Similarly, the 2022 disruption of Black Sea routes highlighted how the blockade of a critical bridge node triggers price volatility that propagates far beyond the immediate trading partners of the conflict zone[27,52].

## 4.3 PageRank and Eigenvector Centrality

To capture the quality rather than the quantity of connections, researchers employ spectral measures such as Eigenvector Centrality and Google's PageRank algorithms[40]. These recursive metrics posit that a node is influential if it is connected to other influential nodes.

High Eigenvector Centrality implies membership in the network's elite core. The United States and China historically dominate this metric, reflecting their centrality within the "Rich Club" of global commerce. A link to the US contributes more to a nation's centrality score than a link to a peripheral state, reflecting the asymmetric weight of establishing trade ties with superpowers.

Recent longitudinal analyses utilizing PageRank ($P$) reveal a structural power shift[1,47,53]. While Western economies remain dominant, there is a statistically significant rise in the centrality of emerging economies, specifically the BRICS nations (Brazil, India). These nations are not merely increasing their volume (Strength) but are increasingly acting as hubs for other influential nodes, effectively decentralizing the historical core-periphery structure of the 20th-century food system.

# 5 Community Structure and Geopolitical Regionalization

While macro-scale metrics (degree distributions) reveal global properties, and micro-scale metrics (centrality) identify individual influencers, the mesoscale architecture of the Global Food Trade Network (FTN) reveals the distinct trading blocs that govern the flow of resources[22]. Through the application of community detection algorithms—most notably Modularity Maximization (e.g., the Louvain algorithm)—researchers can partition the network into functional modules[22,43,54].

Mathematically, this process seeks to maximize the modularity index $Q$, defined as:

$Q = \frac{1}{2m} \sum_{i,j} \left[ A_{ij} - \frac{k_i k_j}{2m} \right] \delta(c_i, c_j)$, where $A_{ij}$ is the weight of the link between $i$ and $j$, $k_i$ is the degree of node $i$, $m$ is the total weight of the network, and $\delta(c_i, c_j)$ is the Kronecker delta function, which is 1 if nodes $i$ and $j$ belong to the same community, and 0 otherwise.

These communities ($c$) represent groups of nations where the density of internal trade significantly exceeds the density of external trade. In the context of the FTN, these mathematical partitions serve as proxies for the de facto geopolitical spheres of influence, often aligning with—but occasionally transcending—geographical continents and formal alliances[19,45].

## 5.1 Taxonomy of Major Trading Modules

Longitudinal analyses reveal a relatively stable partition of the global system into 4 to 7 persistent communities over the last three decades[1,18]. However, the internal topology and leadership dynamics of these blocs have undergone significant evolution.

Historically the most robust community, this module is anchored by the hegemony of the United States. It encompasses the USMCA bloc (Canada, Mexico) and extends deeply into Central America and the Caribbean.

The internal structure exhibits a distinct "Star" topology[55]. The US acts as the central hub, managing high-volume radial flows to peripheral partners. Horizontal connectivity between peripheral members (e.g., direct trade between Caribbean nations) is sparse compared to their vertical integration with the US hub.

Centered around the triad of France, Germany, and the Netherlands, this community represents the operational reality of the European Single Market. In stark contrast to the Americas, the European module resembles a clique or a near-complete graph. High link density exists between almost all members, reflecting the free movement of goods and the Common Agricultural Policy (CAP). This module frequently incorporates former colonial states in West and North Africa, maintaining historical "path dependencies" where trade routes follow 20th-century colonial lines[56].

Emerging from the fragmentation of the post-Soviet space in the 1990s, a distinct "Black Sea Community" solidified post-2010. Dominated by Russia, this bloc includes Ukraine (historically) and Kazakhstan. It functions as a specialized "wheat export machine," serving a cluster of dependent importers primarily in the Middle East and North Africa (MENA), such as Egypt, Turkey, and Iran. This community is structurally fragile. It relies heavily on a limited number of maritime chokepoints (e.g., the Turkish Straits) and is subject to high geopolitical volatility, as evidenced by the trade reconfigurations following the 2022 Russo-Ukrainian conflict[52,56,57].

Often referred to as the "rice bowl" of the global system, this community serves the vast populations of Asia, Oceania, and increasingly Sub-Saharan Africa. Unlike the monocentric American or Russian blocs, this module is polycentric (dual-core). It is typically anchored by Thailand and India, with Vietnam emerging as a third pole. The structure is fluid and dynamic, often reconfiguring annually based on the stochastic nature of monsoon-driven rice harvests.

## 5.2 Compartmentalization Paradox

A critical finding in recent literature is the secular rise of the modularity index $Q$, which has climbed steadily to reach values approximating $Q \approx 0.43$ (circa 2018-2022)[24,28]. This trajectory suggests that despite the rhetoric of "globalization," the food system is becoming increasingly regionalized and compartmentalized. This structural trend presents a fundamental paradox for systemic stability.

High modularity acts as a damping mechanism for diffusion processes[58]. In the event of a biological shock (e.g., a localized pathogen like Xylella fastidiosa or African Swine Fever), strong community boundaries can prevent the contagion from propagating globally[59]. The module serves as a "quarantine zone," sacrificing local stability to protect the global whole.

Conversely, high modularity implies that peripheral nations are structurally "locked in" to their regional hegemon. A peripheral nation in a highly modular community lacks alternative trade partners (low bridging centrality)[60]. Consequently, if the regional hub fails—due to a climate shock or political collapse—the dependent nodes face immediate and catastrophic supply shortages, lacking the connectivity to access "emergency" flows from other communities.

# 6 Spatiotemporal Evolution of Network

The trajectory of the Global Food Trade Network (FTN) is characterized by non-stationarity. It does not follow a linear path of accumulation but rather exhibits distinct phases of "punctuated equilibrium," reflecting the turbulent interplay between global macroeconomic integration and geopolitical fragmentation[61,62]. Longitudinal analysis over the past three decades allows for the demarcation of three specific topological regimes.

## 6.1 Era of Hyper-Globalization (1992–2008)

The post-Cold War era, catalyzed by the dissolution of the Soviet Union and the ratification of the World Trade Organization (WTO) in 1995, marked a period of explosive topological expansion. This era was defined by rapid network densification. The average degree $\langle k \rangle$ of the network surged as nations dismantled tariff barriers and integrated into the global lattice. This period prioritized comparative advantage and logistical efficiency above all other metrics. Structurally, the network exhibited a distinct "unipolar" organization[23,63]. The United States functioned as the undisputed central node—the "hyper-hub"—coordinating global grain flows. Paradoxically, while global connectivity increased, the modularity index ($Q$) also rose. This phenomenon, often termed "glocalization," indicates that as the world integrated, it simultaneously crystallized into distinct regional trading blocs (e.g., NAFTA, EU), laying the groundwork for a hierarchical, community-based architecture[63-65].

## 6.2 "Great Trade Collapse" and Slowbalization (2008–2016)

The 2008 Global Financial Crisis and the concurrent global food price spikes marked a critical phase transition. The exponential growth in trade volume arrested, and the system entered a regime of "Slowbalization"—a period where trade intensity plateaued relative to GDP. The most significant structural shift was the decay of unipolarity. The relative centrality of the United States waned as the network reconfigured around emerging agricultural heavyweights[66,67]. Two major clusters rose to prominence: the "Southern Cone" complex (led by Brazil's dominance in soy and animal protein) and the "Black Sea" complex (driven by the agricultural resurgence of Russia, Ukraine, and Kazakhstan)[2]. This rebalanced the FTN from a star topology into a multipolar system with competing centers of gravity, effectively redistributing the network's "load" and altering global dependency patterns.

## 6.3 Disruption and Resilience (2017–2024)

The current epoch is defined by high-frequency, high-impact exogenous shocks: the Sino-US trade war (2018), the COVID-19 pandemic logistic crunch (2020), and the Russo-Ukrainian conflict (2022). Network analysis reveals a fundamental strategic inversion: a transition from "efficiency-seeking" (cost minimization) to "stability-seeking" (risk mitigation) behaviors. Importers, exposed to the volatility of just-in-time supply chains, have actively sought to increase topological redundancy, deliberately forming trade links that may be economically suboptimal but offer diversification benefits[14,26,27]. Despite the severity of these shocks, the FTN has demonstrated remarkable plasticity—the capacity to dynamically rewire connections in response to node or edge removal. The severance of direct flows between Russia and the EU did not result in a global void; rather, it triggered a rapid rearrangement of flux. Russian exports pivoted toward Asian and African markets, while Europe substituted supply via the Americas. While this adaptive rewiring prevented systemic collapse, it introduced higher friction. The new network configuration is characterized by longer average path lengths for certain commodities and increased transaction costs, effectively functioning as an "insurance premium" paid for global food security.

# 7 Crop-Specific Topologies: The Multiplex of Commodities

It is a methodological reductionism to treat the Global Food Trade Network (FTN) as a monolith. In reality, the FTN operates as a Multiplex (or Multilayer) Network, denoted formally as $\mathcal{M} = \{G_\alpha\}_{\alpha=1}^{L}$, where each layer $\alpha$ corresponds to a distinct commodity (e.g., wheat, rice, soy) with its own unique topology $G_\alpha = (N, E_\alpha, W_\alpha)$. Aggregating these layers into a single unweighted graph obscures critical structural heterogeneities; the architecture of food security varies profoundly depending on the specific caloric vehicle in question[67-69].

## 7.1 Wheat: The Decentralized Backbone

The wheat layer ($G_{wheat}$) represents the most globalized and topologically robust component of the system[67,70,71]. Due to the crop's agronomic versatility across diverse latitudes, production is distributed among a wide array of active exporters (the United States, Canada, Russia, France, Australia, Argentina). This geographical dispersion creates a network with high source redundancy. The topology typically exhibits high clustering and multiple independent pathways between communities. Consequently, the wheat network demonstrates significant robustness to localized supply shocks; a harvest failure in one breadbasket can largely be compensated by surplus in another, buffering the system against total collapse, although it remains sensitive to price transmission from the critical Black Sea axis.

## 7.2 Rice: Regional Fragility and the "Single Point of Failure"

In sharp contrast, the rice layer ($G_{rice}$) is characterized by extreme regionalization and structural brittleness. The network is heavily centered on the Asian monsoon belt. The export market is oligopolistic, dominated by a "troika" of suppliers: India, Thailand, and Vietnam. The emergence of India as the hyper-dominant hub (ranking 1st in out-degree and out-strength) has introduced a systemic "Single Point of Failure" risk. The topology lacks the redundancy of wheat; therefore, policy-induced shocks—such as India's 2023 export ban—do not merely re-route flows but effectively sever the supply lines for dependent importers in Sub-Saharan Africa and Southeast Asia[72]. The rice network is structurally "brittle"—highly efficient under normal conditions but prone to catastrophic fragmentation under stress[73,74].

## 7.3 Soybeans: Polarized "Dumbbell" Topology

The soybean layer ($G_{soy}$) exhibits the most distinct and potentially precarious architecture, defined by extreme bipartite concentration. On the supply side, the network is a rigid duopoly controlled by the Americas (Brazil and the USA). On the demand side, it forms a near-monopsony centered on China[36,67,70]. The resulting topology resembles a massive "dumbbell" or polarized structure, with thick, high-volume arteries connecting the Western Hemisphere to East Asia. While logistically hyper-efficient, this structure possesses near-zero redundancy. A climatic failure in Brazil or a geopolitical rupture between the US and China leaves the system with no alternative rewireable pathways. This layer is the primary vector for the transmission of trade-war shocks into the global food system[75].

## 7.4 Maize: Feed-Food Nexus

The maize network ($G_{maize}$) occupies an intermediate topological position, serving as the critical bridge between the food system and the global livestock/ethanol industries. While sharing the American-centric hubs of the soybean network, the maize layer historically relied on a distinct "European connector": Ukraine. Ukraine's role as a high-betweenness bridge node—linking Eastern production to EU and Chinese consumption—was a defining feature. The disruption of this node has forced a rewiring of the network, increasing the load on the remaining American hubs and tightening the coupling between energy markets (biofuels) and food security[2,76].

## 7.5 Beef: Shifting Center of Gravity

Longitudinal analysis of the beef trade layer ($G_{beef}$) reveals a dynamic "core-shift" reflecting the global dietary transition. While the supply core remains anchored in the Americas and Oceania (Brazil, Australia), the demand core has migrated rapidly from the Russian Federation and Western Europe to East Asia (China, Vietnam). This layer is currently undergoing a phase of densification (increasing average degree $\langle k \rangle$) as rising incomes in the Global South drive a protein transition[77-80]. The network is becoming increasingly connected, yet it shows distinct signs of clustering, implying that the "meatification" of global diets is creating new, regionally distinct trade dependencies that differ from traditional grain corridors.

# 8 Multiplex Frontier: Food, Energy, and Virtual Water

Contemporary network science increasingly rejects the reductionist view of the food system as an isolated topological layer. Instead, it is conceptualized as a constituent layer within a Multiplex Network or a "Network of Networks" (NON), intimately entangled with global energy fluxes and hydrological resources.[2,68,69] This framework, known as the Food-Energy-Water (FEW) Nexus, posits that a node (nation) $i$ operates simultaneously across multiple interconnected layers: the food trade layer ($G_{food}$), the energy trade layer ($G_{energy}$), and the virtual water trade layer ($G_{water}$). Understanding systemic risk requires analyzing the coupling strength and inter-layer dependency between these systems.

## 8.1 Virtual Water Flows

The concept of "virtual water"—defined as the volume of freshwater embodied in the production chain of a commodity—transforms the FTN into a mechanism for the anthropogenic redistribution of global water resources.

Network analysis reveals that the VWTN functions as an efficiency engine. By facilitating the flow of water-intensive crops (e.g., cereals) from nations with high water productivity (temperate, rain-fed zones like North America and Northern Europe) to nations with low water productivity (arid, irrigation-dependent zones), the network generates a net "global water saving."[36]

The topology of the VWTN, while structurally isomorphic to the FTN, highlights distinct vulnerabilities[81]. It reveals that food security in the MENA region (Middle East and North Africa) is fundamentally a problem of hydrological subsidization. These nations do not merely import wheat; they import the water required to grow it. Consequently, a disruption in the grain supply from the Black Sea is not just a caloric shock; it is effectively an acute water crisis for arid importers who lack the "green water" reserves to substitute domestic production[71,81].

## 8.2 Energy-Food Coupling

The interdependence between the energy and food layers represents the most volatile frontier of systemic risk. This coupling is mediated primarily through input costs: energy prices drive the cost of feedstock for fertilizer production (specifically natural gas for the Haber-Bosch process) and logistical transport[2].

Shocks in the energy layer ($G_{energy}$) propagate to the food layer ($G_{food}$) with time-lagged but amplified intensity[67]. High hydrocarbon prices degrade the terms of trade for agricultural producers, forcing a reduction in yield or an increase in export prices.

The Russia-Ukraine conflict serves as the quintessential case study of a multiplex cascading failure[82-85]. It was not a singular crisis but a synchronous disruption across three coupled layers: In Energy Layer, sanctions and supply cuts in natural gas spiked production costs. In Fertilizer Layer, the "hidden centrality" of Belarus and Russia—who command a dominant share of the global potash and nitrogen markets—acted as a contagion mechanism. Restrictions on these nodes throttled global fertilizer availability. In Food Layer, this supply shock transmitted inflation to distal nodes like Brazil (a major fertilizer importer), which, faced with rising input costs, transmitted price shocks back to the global market via soybean and maize exports.

# 9 Resilience, Vulnerability, and Impact of Shocks

In the lexicon of complex systems science, resilience is formally defined as the capacity of the Global Food Trade Network (FTN) to absorb stochastic perturbations—whether climatic extremes, geopolitical ruptures, or economic volatility—and maintain its critical functionality: the provisioning of caloric security[86-88]. Conversely, vulnerability denotes the system's susceptibility to state degradation or topological fragmentation under stress. Recent empirical literature highlights a fundamental structural tension governing these properties: the dialectic between economic optimization and systemic robustness[89-91].

## 9.1 Efficiency-Resilience Trade-off

Network architecture is governed by two competing optimization functions:

An efficiency-maximizing topology concentrates production in regions with the highest comparative advantage (e.g., the hyper-concentration of soy production in the Brazilian Cerrado) and consolidates flows into high-volume arteries to exploit economies of scale. Structurally, this results in a sparse, centralized network characterized by low redundancy and "Just-in-Time" logistics[92,93].

A resilience-maximizing topology prioritizes redundancy[94]. It maintains a diversity of supply nodes and alternative pathways, even if these links are economically suboptimal under steady-state conditions. Structurally, this results in a dense, distributed network capable of "Just-in-Case" responsiveness.

Historically (1990–2015), the evolution of the FTN was driven by the imperative of efficiency[23,63]. However, the post-2017 geopolitical landscape has catalyzed a phase transition[95,96]. Network actors are increasingly pricing in a "resilience premium," deliberately diversifying trade partners to buffer against volatility. This represents a structural retreat from the hyper-efficient but brittle architectures of the past toward more robust, albeit costlier, configurations.

## 9.2 Cascading Failures and Shock Propagation

The hyper-connected nature of the FTN implies that local perturbations rarely remain contained; instead, they propagate through the lattice via cascading failures[21]. The mechanism of this propagation is non-linear and often amplified by behavioral feedback loops[62].

A typical cascade follows a predictable trajectory: 1. A stochastic event (e.g., a heatwave in the Pontic Steppe) reduces yield in a core node (e.g., Russia)[97]. 2. To stabilize domestic inflation, the affected core node imposes export restrictions or quotas. This policy response effectively removes the node from the network, severing links[98].

3. Direct trade partners (e.g., Egypt, Turkey) are forced to enter the global spot market to secure replacement volumes[67,71]. This sudden, inelastic demand surge triggers a global price spike. 4. The price signal transmits the shock to distal, import-dependent nodes (e.g., Indonesia, Nigeria) that were topologically disconnected from the original event[99]. These nations, facing fiscal instability, may engage in panic buying or subsidy expansion, further amplifying the volatility[100].

Empirical modeling of the 2008 food price crisis and the onset of the COVID-19 pandemic demonstrates that trade policy is the primary determinant of shock magnitude[101,102]. Cooperative trade policies function as dampeners, distributing the shortfall across the network. In contrast, protectionist responses (export bans/hoarding) act as multipliers, magnifying a manageable local deficit into a systemic global crisis. The network is thus highly sensitive not just to the physics of production failure, but to the game-theoretic behavior of its major hubs[96].

# 10 Impact of Recent External Shocks

The theoretical vulnerabilities of the Global Food Trade Network (FTN) have recently moved from the realm of computational simulation to empirical reality. The system has been subjected to two consecutive, high-magnitude "Black Swan" events—the COVID-19 pandemic and the Russia-Ukraine conflict—which functioned as massive, real-world stress tests[103,104]. These events offer distinct case studies in network dynamics: the former representing a synchronous logistical dampening affecting edges, and the latter representing a targeted node suppression affecting critical hubs.

## 10.1 Russia-Ukraine Conflict

The 2022 conflict constituted a critical disruption to the "Black Sea Granary," a cluster that accounts for a substantial fraction of global exports in wheat, maize, and sunflower oil. From a network perspective, this event was characterized by the physical and financial severance of the edges connecting these core production nodes to the global giant component.

The conflict effectively isolated the "Russian Community" (identified in modularity analyses) from Western markets via sanctions and from the Global South via the physical blockade of Black Sea ports. This disruption served as a test of the network's modular resilience[105].

Despite predictions of systemic collapse, the FTN demonstrated remarkable topological plasticity. The network underwent a rapid, large-scale reconfiguration[62,96]. Russian exports, sanctioned by the G7, reoriented their trajectory toward "friendly" or neutral nodes in Asia and Africa (e.g., China, India, Egypt). Conversely, the European Union, facing a sudden deficit in Ukrainian feed maize and sunflower oil, executed a substitution strategy, rewiring its supply chains to source from the Americas (Brazil, USA)[106,107].

While the structure of the network survived—meaning the global giant component did not fracture—the function was impaired. The cost of this forced rewiring manifested as record-high volatility in the FAO Food

Price Index[108]. The crisis highlighted that while the network is robust to total failure, its adaptability comes at the cost of immense financial friction.

Crucially, the conflict exposed the "hidden centrality" of the fertilizer layer[109]. While Ukraine's centrality in the wheat layer declined (dropping from a top-5 exporter), the disruption of the Russia-Belarus fertilizer node propagated a shockwave into the production cost functions of virtually every other agricultural node in the network, illustrating the dangerous coupling between the food and energy layers.

## 10.2 COVID-19 Pandemic

In contrast to the localized production shock of the war, the COVID-19 pandemic (2020–2022) represented a systemic logistical shock that applied stress to every edge in the graph simultaneously.

Lockdowns, quarantine protocols, and labor shortages in port facilities functioned as an attack on the links (edges) of the network rather than the nodes[110,111]. This resulted in a global increase in "resistance" or friction across the entire topology, slowing the velocity of trade flows.

Contrary to dire forecasts of a trade collapse, empirical data reveals that the total volume of global food trade actually increased during the pandemic[112,113]. This counter-intuitive resilience was driven by the redundancy accumulated over the previous decade. The FTN's "Small-World" architecture allowed suppliers to bypass congested nodes and identify alternative logistical pathways[49,96]. Furthermore, the shock triggered a behavioral response—strategic hoarding and panic buying—which paradoxically stimulated trade volume despite logistical constraints.

The long-term topological legacy of the pandemic is the acceleration of network contraction. The vulnerability of extended, just-in-time supply chains has prompted nations to prioritize "Near-Shoring" and "Friend-Shoring." [114]We are currently observing a structural shift where the average path length of food trade is decreasing as nations seek to shorten supply chains, trading efficiency for the security of regional proximity[115].

# 11 Future Directions: Climate Change and Network Adaptation

While geopolitical shocks are transient, the secular trend of anthropogenic climate change represents a permanent, non-stationary forcing mechanism acting upon the Global Food Trade Network (FTN). Forward-looking network modeling suggests that warming temperatures will not merely alter yield volumes but will fundamentally restructure the topology of global calorie flows[67,97].

## 11.1 Poleward Migration of Comparative Advantage

Agro-climatic modeling projects a significant poleward shift in production potential. As isotherms migrate, the comparative advantage of traditional breadbaskets in tropical and sub-tropical latitudes (e.g., Brazil, India, Sub-Saharan Africa) is projected to decline due to heat stress and hydrological volatility[116]. Conversely, high-latitude nations (principally Russia, Canada, and Northern Europe) may experience a "boreal expansion" of arable frontiers[117]. This dictates a radical rewiring of the global adjacency matrix. The FTN is expected to transition from a multi-polar structure (with production centers in both the Northern and Southern hemispheres) to a increasingly uni-polar or bi-polar structure anchored in the Northern Hemisphere. This reorganization implies that the global South will become increasingly structurally dependent on the global North for caloric sustenance, exacerbating existing asymmetries in the network[96].

## 11.2 Paradox of Hyper-Centralization

A critical finding from recent simulation studies is the risk of systemic consolidation. As climate stress renders marginal producers unviable, the burden of supplying the global buffer stock will concentrate onto a shrinking subset of "mega-exporters" located in climate-resilient zones[96]. This contraction in the number of effective source nodes ($N_{source}$) creates a topological paradox. While the individual capacity of the remaining hubs may increase, the systemic resilience decreases[62,118]. The network moves toward a "Star-like" topology with extreme centralization. If the global food system relies on three breadbaskets instead of five, the failure of a single hub—due to a "Black Swan" event like a concurrent multi-breadbasket failure—becomes a catastrophic systemic event. The redundancy that characterizes the current "Small-World" network is stripped away, leaving a rigid, brittle skeleton that lacks the plasticity to absorb shocks[49,94].

## 11.3 Transboundary Adaptation and the Isolationist Trap

Future resilience relies on the concept of "Adaptation without Borders"—the recognition that climate risk is teleconnected. A drought in the US Midwest is not a local hazard; via the FTN, it is a transboundary risk for Egypt and Japan[118]. Network theory warns that the intuitive policy response to scarcity—isolationism and export restrictions—constitutes a "Nash Equilibrium" that is disastrous for the system as a whole[98,101]. In a highly coupled, climate-stressed system, protectionist cascades (where node A hoards, forcing node B to panic) amplify volatility[113,119]. The mathematical imperative for the next decade is the design of cooperative trade protocols that prevent these cascading failures. Resilience cannot be achieved by individual nodes optimizing for local security; it requires a governance architecture that maintains open edges even when nodal stress is high.

# 12 Conclusion

The application of complex network theory to global food trade has fundamentally transformed our understanding of food security. The FTN is revealed not merely as a collection of bilateral commercial agreements, but as a dynamic, evolving complex system with distinct topological properties. It is a "small-world" network that efficiently connects distant producers and consumers, yet it is also a "scale-free" system with a pronounced hierarchy of hubs and peripheries.

The historical trajectory of the network—from the unipolar, efficiency-obsessed era of the 1990s to the multipolar, fragmented, and resilience-focused era of the 2020s—reflects a defensive adaptation to a volatile world. The rise of regional trading blocs and the increasing modularity of the system serve as containment mechanisms against systemic risk. However, as the Russia-Ukraine conflict and the COVID-19 pandemic have demonstrated, the system remains prone to cascading failures, particularly when key "bridge" nodes or "hub" exporters are disrupted.

Future policy and research must move beyond the binary of "trade vs. self-sufficiency" and embrace the concept of "network resilience." This requires strategies that enhance diversity, promote redundancy, and foster international cooperation to keep the edges of the network open. In an era of climate instability, the resilience of the Global Food Trade Network is not just an economic concern; it is a prerequisite for the survival of modern civilization.

# Acknowledgements

This paper is supported by Zhejiang Provincial Philosophy and Social Sciences Planning Project (Grant No. 24NDJC175YB). C.T. acknowledges support from the National Natural Science Foundation of China (Grant No. 72571247) and Scientific Research Project of Zhejiang Provincial Bureau of Statistics (Grant No. 25TJZZ18).

# Conflicts of Interest

The authors declare no conflicts of interest.